# Correcting Doppler Shifts in He II 30.38 nm Line by Using the EVE and AIA data from Solar Dynamics Observatory

Zhixun Cheng[1,2], Yuming Wang[1,2,*], and Rui Liu[1,2]

[1] CAS Key Laboratory of Geospace Environment, School of Earth and Space Sciences, University of Science and Technology of China (USTC), Hefei, 230026, China

[2] CAS Center for Excellence in Comparative Planetology, USTC, Hefei, 230026, China

*Corresponding author, E-Mail: ymwang@ustc.edu.cn

## Abstract

The Extreme-ultraviolet Variability Experiment (EVE) onboard the Solar Dynamics Observatory (SDO) detects the solar EUV spectra with high temporal cadence and spectral resolution. The wavelength shifts of emission lines provide key information of the dynamics of the Sun. However, some of EVE spectral observations are influenced by the non-uniformly distributed irradiance on the Sun, which may prevent us from correctly understanding the physical processes happened in the solar corona. Here, based on the only published on-orbit calibration data of EVE He II 30.38 nm line on 27 Jan 2011 (Chamberlin, 2016), we develop a method to correct the He II 30.38 nm line by using AIA 304 imaging data. This correction method is then applied to EVE He II 30.38 nm data from 29 Oct 2010 to 3 Mar 2011 to study the Doppler oscillations of the solar He II 30.38 nm line, in which we show that the half-month periodic Doppler oscillation is caused by non-uniformly distributed irradiance mainly due to the presence of active regions. Other EVE coronal lines also present similar Doppler oscillations, suggesting that an appropriate correction must be implemented before interpret the oscillation phenomena appearing in these lines.

1. Introduction

The main target of The Extreme-ultraviolet Variability Experiment (EVE; Woods et al., 2012) onboard the Solar Dynamics Observatory (SDO; Pesnell et al., 2012) is to detect EUV irradiance variations of the Sun viewed as a star, but its sensitivity and stability of spectral measurement suggest a capability to study wavelength shifts of emission lines. Some previous papers found that EVE can in fact measure wavelength shifts during solar eruptive events and therefore derive Doppler velocities for plasma at all temperatures from the chromospheres to hot flaring regions. Hudson et al. (2011) first demonstrated the possibility of dynamic measurement on flaring plasma based on EVE spectrum by measuring the Doppler shifts of He II 30.38 nm line and Fe XXIV 19.2 nm line. Brown et al. (2016) reported the plasma speeds in EVE hydrogen Lyman and C III line during several M and X class flares. Recently, Cheng et al. (2019) provided a comprehensive dynamic analysis on flares by investigating 10 EVE emission lines which cover the temperature range from the transition region to the core of flare.

Besides the short-term measurements on solar eruptive events, the continuous observations of EVE also give us an opportunity to detect the long-term evolution of plasma Doppler velocities of the global Sun. Such long-term detection for plasma velocity would help to construct a fundamental dynamic pattern of solar atmosphere, e. g., initial acceleration of solar wind, emergence and disappearance of active regions and modulation of solar cycles.

However, one of the challenges in EVE long-term Doppler measurements is the calibration of central wavelength of EUV lines. The Multiple EUV Grating Spectrograph A channel (MEGS-A), one of the three EVE spectrograph channels, is a grazing incident spectrograph and is sensitive to any off-axis shifts (Crotser et. al, 2007). Not like light from the center of the visible solar disk, the light from the limb of the Sun has a slight offset in wavelength. This optical characteristic is not a problem for irradiance measurement, as it only makes the full-disk irradiance line profile slightly broader. But it does matter when determining the central wavelength of the line profile. This instrumental effect does not need to be considered if just observing relative Doppler velocities during different stages of an eruptive event as was done in Hudson et al. (2011) and Cheng et al. (2019), which compared the relative velocities between the impulsive and gradual phases. But such optical offset must be estimated and subtracted for the study of the long-term Doppler variation of the Sun.

Hudson et al. (2011) pointed out this type of optical offset of EVE MEGS-A and gave a correction equation for the inherent optical shift as a function of event location on the solar disk. Chamberlin (2016) subsequently presented a complete calibration analysis for He II 30.38 nm line by using on-orbit "cruciform scans", and provided a more accurate correction formula for the emission line wavelength. But EVE data have no spatial resolution and the on-orbit cruciform scans were not performed continually. Thus, the non-uniform and time-changing distribution of the emission intensity on the solar disk will cause the instrumental effect hard to be removed from the EVE data without the aid of other imaging data, e.g., AIA images. In this paper, we move step forward to establish a method to remove the instrumental effect in wavelength based on the correction formula obtained from the cruciform scans and the SDO/AIA images. This method will make the long-term analysis of the Doppler variation of the global Sun possible.

## 2. Method
### 2.1. MEGS-A Cruciform Calibration

Cruciform scans were performed by slightly adjusting the pointing direction of the SDO spacecraft in order to obtain irradiance line profiles with the optical shifts at different offset angles. The two offset axes in the scans are along the latitude and longitude directions of the Sun, corresponding to the dispersion and cross-dispersion axes of the MEGS-A channel, respectively.

Chamberlin (2016) investigated He II 30.38 nm line with the cruciform scans and derived correction equation for the instrumental wavelength offset of this line by studying the relationship between wavelength shifts and offset angles. The correction as a function of offset angle is found to be well fitted as a linear function for latitude offset and a quadratic function for longitude offset. Therefore, the latitude offset can cause both red- and blue-shift of the line center, while the optical wavelength shift in longitude is always red-shift.

The wavelength shift for He II 30.38 nm line on 27 Jan 2011 is given as (Chamberlin, 2016)

$$\Delta\lambda = 0.0752\alpha^2 + 0.265\beta, \tag{1}$$

or

$$\Delta\lambda = 19.8\sin^2\varphi + 4.3\sin\theta, \tag{2}$$

in which $\Delta\lambda$ is in units of pm, and $\alpha$ represents the offset angle along dispersion direction of the MEGS-A grating and $\varphi$ is the corresponding longitude on the solar disk, where east longitudes are input as negative and west longitudes are positive. Similarly, $\beta$ represents the offset angle along the cross-dispersion direction of MEGS-A and $\theta$ is the corresponding latitude with north latitude negative and south positive. The unit of $\alpha$ and $\beta$ is arcmin in Equation (1) and (2). A corrected wavelength can be derived from the measured wavelength by subtracting the wavelength shift, i.e., $\lambda = \lambda_m - \Delta\lambda$.

## 2.2. Simulation based on the AIA imaging data

The correction method in Chamberlin (2016) regards the Sun as a star when performing the cruciform scans, since EVE has no spatial resolution. Here we present a simulation based on AIA imaging data to estimate the influence from the spatial size and disk irradiance distribution of the Sun.

He II 30.38 nm is designed to be the primary ion of AIA 304Å passband (Lemen et al., 2012), and O'Dwyer et al. (2010) suggest that He II contributes 97% of AIA 304 emission in quiet Sun and more than 80% in active regions. So we select the AIA 304 image at 22:58:56 UT on 27 Jan 2011 as the original data, as shown in Figure 1(a), which was observed just after the cruciform scan on the same day performed by Chamberlin (2016). For every pixel in the image, we construct a Gaussian line profile to represent the He II 30.38 nm irradiance coming from that pixel as shown in Figure 1(b). The wavelength shift of line center is derived from Equation (2), while the value of line peak is determined by the observed AIA 304 intensity of the pixel. Because the full width at half maximum (FWHM) of the EVE He II 30.38 nm line is about 0.1 nm, the simulated line profile for each AIA 304 image pixel is set to have 0.1 nm FWHM, by assuming that the AIA 304 channel has similar Doppler temperatures across the entire solar disk.

Then we integrate the results of all pixels in the AIA image and obtain the full-disk irradiance profile without SDO's offset motion, as shown in Figure 1(c). The full-disk profile obeys a single-peak distribution, slightly deviating from Gaussian distribution due to the EVE's optical offset. Meanwhile, the central wavelength of the profile shows a small red-shift because of the spatial size of the Sun. According to Equation (2), the east and west hemispheres will both produce red-shift optical offset, though the blue-shift from the north hemisphere and red-shift from the south hemisphere may tend to cancel each other.

In order to simulate the cruciform scans of EVE, we adjust the center of the AIA image to make a certain offset angle, use Equation (1) to generate line profile for each pixel, and then add up the irradiance profiles of all pixels again for a full-disk irradiance distribution containing offset effect of SDO. This procedure is repeated at different offset angles ranging from -30' to 30', with an interval of 1', in both the latitude and longitude directions, respectively. Figure 2(a) and 2(b) shows the full-disk irradiance profiles at different offset angles along the alpha and beta offset axes, i.e., longitude and latitude directions. Figure 2(a) indicates the simulated He II 30.38 nm profiles for alpha offsets from -30' to 0' with significant red-shifts, while the situation from 0' to 30' is similar and therefore not plotted in the figure for a clear view. Figure 2(b) indicates the profiles for beta offsets from -30' to 30' with slight blue- and red-shifts.

Discrete data points are selected in the irradiance curves with the EVE spectral interval of 0.02 nm and are then used for a single Gaussian fit to determine the central wavelengths of the full-disk profiles and therefore the corresponding optical shifts. Figure 2(c) and 2(d) shows the wavelength shifts as a function of offset angle along alpha and beta axes. The red symbols and solid curves represent the simulation and fitting results, and the black dashed curves are the EVE observation during the cruciform scan on 27 Jan 2011 in Chamberlin (2016), which is described by Equation (2). The zero point on the y-axis is set to be the rest wavelength without any SDO offset, i.e., $\alpha$ and $\beta = 0'$. The alpha offset effect is well fitted by a quadratic function, which basically matches the EVE cruciform scan observation. At the east and west limbs of the Sun, where $\alpha = \pm 16'$ (shown by the vertical dotted lines), the alpha offset causes an optical shift of 20 pm, corresponding to a Doppler velocity of 200 km s$^{-1}$ for He II 30.38 nm line. The difference of correction curve between our simulation and actual cruciform scan observation at $\alpha = \pm 16'$ is 1.5 pm, i.e., 15 km s$^{-1}$ as Doppler velocity. Meanwhile, the beta offset effect obeys a linear function, completely coinciding with the EVE cruciform scan observation, and causes 4 pm shift at the north and south limbs, which is much smaller than the alpha offset. The simulation from the AIA 304 image roughly conforms to the EVE cruciform scan observation of He II 30.38 nm line, suggesting that the correction formula by Chamberlin (2016) is also applicable to the AIA images.

## 2.3. Improvement of Correction Formula

The correction formula by Chamberlin (2016), i.e., Equation (2), is obtained by deviating the whole Sun from the center of EVE's field of view with particular angles. This procedure actually calibrates the optical characteristics of EVE with an extended light source instead of a point light source, which might lead to slight deviation when applying the correction formula directly to a local structure, like an active region, on the solar disk. Especially, the EVE's cruciform scan method probably underestimates the instrumental red-shift offset in alpha direction when ignoring the spatial size of solar disk and regarding the Sun as a star. Unlike the situation in the north and south hemispheres, the red-shift offsets caused by the east and west solar hemispheres obey a quadratic relationship and thus cannot cancel each other to make the Sun equivalent to a point light source. So the alpha correction coefficient derived from cruciform scans would be inaccurate. That is also why the AIA simulation does not completely conform to the EVE cruciform scan observation in alpha direction, as shown in Figure 2(c).

Therefore we adjust the alpha correction coefficient, A, which is 19.8 in Equation (2), and calculate new correction curves repeatedly to find a more accurate correction formula. The mean-square error (MSE) between the correction curves of AIA simulation and EVE observation in alpha direction is used to determine the best value of A. Figure 2(e) shows the relationship between MSE and A. The value of MSE reaches its minimum when A equals to 18.66, resulting a new correction formula for He II 30.38 nm line on 27 Jan 2011

$$\Delta\lambda = 18.66\sin^2\varphi + 4.3\sin\theta, \qquad (3)$$

where $\Delta\lambda$ is in units of pm, and $\varphi$ and $\theta$ are the longitude and latitude of solar event on the solar disk, respectively. The new correction curve of alpha offset is shown in Figure 2(f), which is well consistent with the actual observation from cruciform scans.

It should be noted that this correction formula is only applicable for He II 30.38 nm line within a short time period near the cruciform scan on 27 Jan 2011, because the coefficients in Equation (3) change with time due to the non-uniform degradation across the CCD of EVE (Chamberlin, 2016). The degradation is a function of the amount of exposure and is thus more significant at line center compared to that in the wings. This effect could change the detected line profile mainly by weakening the line intensity and making the profile flatter, especially in the later part of EVE's life. But the non-uniform degradation itself would hardly produce any wavelength shift, as it should occur symmetrically on both wings of a line. Regular cruciform calibrations are preformed every four months over the SDO mission. In order to obtain a time-dependent correction formula of line wavelength, we can apply above method to each cruciform scan by using the corresponding AIA images.

## 3. Doppler Oscillations of some EVE lines
### 3.1. Correction for Long-term EVE He II 30.38 nm Observations

In the following, we present the correction result of He II 30.38 nm line by using Equation (3) over four months around 27 Jan 2011. Figure 3(a) shows the temporal evolution of original and corrected He II 30.38 nm Doppler velocity from 29 Oct

2010 to 3 Mar 2011, with a time cadence of one day, as well as the corresponding correction based on the AIA 304 images at 00:04 UT on each day. EVE MEGS-A provides spectral observations at 8640 times every day. We use Gaussian fitting technique (Cheng et al., 2019) to obtain the central wavelength of the He II 30.38 nm line for these 8640 times. Their median value is used as the central wavelength on that day. By setting the central wavelength on 3 Feb 2011, when the only active region was close to the solar disk center and would not produce apparent optical offset of EVE, to be the reference wavelength, i.e., zero shift, we calculate the Doppler velocity for the whole period from 29 Oct 2010 to 3 Mar 2011, as shown by the solid line in Figure 3(a). The integration of optical offsets from all pixels in the AIA 304 image gives the wavelength correction value on each day, as described in Section 2.2. Original Doppler velocity of He II 30.38 nm and the AIA correction have consistent oscillation patterns with amplitudes of about 10 km s$^{-1}$, suggesting that the wavelength measurement of He II 30.38 nm is indeed influenced by the optical offset in EVE MEGS-A. Such oscillations of optical offset are probably caused by the rotation of active regions, which result in the change of irradiance distribution on the solar disk. Corrected Doppler velocity of He II 30.38 nm line is then calculated by subtracting the AIA 304 correction from the original observations, as shown by the dashed curve in Figure 3(a). Corrected Doppler velocity has similar amplitudes with the original observation but shows a different oscillation pattern. Besides, the Doppler velocity converts to slight blue-shift after the correction, which is not very reliable in the physical sense considering that the determination of absolute Doppler velocity strongly depends on the selection of the reference wavelength.

Fast Fourier Transforms (FFTs) are applied to both original and corrected Doppler velocity sequences of He II 30.38 nm over the four months, and their main oscillation periods are shown in Figure 3(b). There is a significant peak at 14 days in the original velocity period, which is close to half of the solar rotation period. This 14-day period can be interpreted by the rotation of active regions across the solar disk. According to Equation (3), when an active region moves from the east limb to the west limb of the Sun, its optical effect of red-shift will start from a maximum at the east limb, then become minimum near the disk center, and finally reach another maximum at the west limb. Such a half solar rotation causes a complete oscillation of the optical red-shift, as shown in Figure 3(c) - (e). There was a main active region close to the solar disk center on 3 Feb 2011, corresponding to a minimum red-shift. On 27 Jan 2011 and 9 Feb 2011, several active regions were located near the solar limbs due to the solar rotation, which cause the maximums of optical red-shift. If there are several active regions on the visible solar disk, their integrated effect also follows the above rule. This 14-day period is no longer obvious in the corrected velocity, as shown by the dashed line in Figure 3(b), indicating that our wavelength correction procedure effectively reduces the influence of the instrumental offset of EVE. Another small peak at 9 days in the original data disappearing after the correction is due to the same reason. The main period of 31.5 days is unaffected by the wavelength correction and probably reflects the 27-day solar rotation and/or other real physical effect from some dynamic processes on the Sun.

### 3.2. **Long-term observations of other hot coronal lines**

We further study the Doppler velocities of some hot coronal lines from EVE MEGS-A to see whether and how the optical offset of the instrument impacts other emission lines. Long-term observations of Fe XVI 33.54 nm, Fe XV 28.42 nm, Fe XII 19.51 nm, Fe XI 18.04 nm and Fe IX 17.11 nm from 29 Oct 2010 to 3 Mar 2011 are presented in Figure 4, respectively, as well as the AIA 304 correction profile and their correlations. Most coronal lines are found to have high correlation coefficients (CC > 0.58) with the optical correction, except for Fe IX 17.11 nm for which the value of CC is 0.31, with oscillations of half-month period which are similar to the case of He II 30.38 nm. As comparison, the correlation coefficient for He II 30.38 nm is 0.54. This phenomenon indicates that other EVE emission lines are also under the influence of the instrumental factors. Meanwhile, the hotter lines have greater amplitudes of Doppler velocity and better correlations with the optical correction and therefore seem to be easier affected by rotating active regions on the Sun. This result is natural considering the fact that most high-temperature emission comes from active regions instead of background corona.

## 4. Conclusion

In this paper, we improve the correction method of EVE's optical offset effect given by Chamberlin (2016) by combining the AIA 304 image. The spatial size of the Sun and non-uniformity on the solar disk, e.g., active regions, are taken into account through investigating every AIA pixel across solar disk. The correction formula is adjusted for a better description of EVE's optical offset features.

Then we apply the correction method to the long-term measurement of He II 30.38 nm by using EVE and AIA data. The consistence between He II Doppler velocity and AIA 304 correction suggests that the wavelength measurement of He II 30.38 nm is indeed affected by the optical offset of EVE. The 14-day and 9-day periods of Doppler oscillation can be interpreted by the rotating active regions across the solar disk, which disappears after our wavelength correction procedure. Other EVE coronal lines present similar Doppler oscillations as He II 30.38, indicating the optical offset of EVE significantly impact emission lines of different wavelengths and should be removed before doing any long-term analysis on these lines.

The future work could focus on the determination of time-dependent correction formulas of He II 30.38 nm and other EVE lines by combining the actual observation of EVE's cruciform scans and the corresponding AIA imaging data. If the wavelength corrections of all EVE lines were done, we could obtain real Doppler oscillations at multiple wavelengths, which will help us understand the long-term dynamic evolution of the Sun. Due to the importance of long-term analysis of the EVE spectral lines, we strongly suggest that the benchmark work, such as the calibration of the EVE cruciform scans data, should be supported.

We acknowledge use of the EVE and AIA data from the SDO spacecraft. SDO is a mission of NASA's Living With a Star Program. The EVE version 6 data are used and can be retrieved at https://lasp.colorado.edu/eve/data_access/. The AIA imaging data are obtained from https://sdo.gsfc.nasa.gov/data/aiahmi/. We thank the anonymous reviewer for providing constructive comments, and we are grateful to EVE team for help in improving the paper and addressing some of the comments.

This work is supported by the Strategic Priority Program of the Chinese Academy of Sciences (No. XDB41000000) and the NSFC (No. 41774178).

**References**

Brown, S.A., Fletcher, L., Labrosse, N., Doppler speeds of the hydrogen Lyman lines in solar flares from EVE, 2016, A&A, 596, A51.

Chamberlin, P.C., Measuring solar Doppler velocities in the He II 30.38 nm emission using the EUV Variability Experiment (EVE), 2016, Solar Phys. 291, 1665.

Cheng, Z., et al., Plasma motion inside flaring regions revealed by Doppler shift information from SDO/EVE observations, 2019, ApJ., 875, 93.

Crotser, D.A., et al., SDO-EVE EUV spectrograph optical design and performance, 2007, Society of Photo-Optical Instrumentation Engineers (SPIE) Conference Series, Society of Photo-Optical Instrumentation Engineers (SPIE) Conference Series, 6689.

Hudson, H.S., et al., The EVE Doppler sensitivity and flare observations, 2011, Solar Phys., 273, 69.

Lemen, J. R., et al., The Atmospheric Imaging Assembly (AIA) on the Solar Dynamics Observatory (SDO), 2012, Solar Phys., 275, 17.

O'Dwyer, B., et al., SDO/AIA response to coronal hole, quiet Sun, active region, and flare plasma, 2010, A&A, 521, A21.

Pesnell, W.D., Thompson, B.J., Chamberlin, P.C., The Solar Dynamics Observatory (SDO), 2012, Solar Phys. 275, 3.

Woods, T.N., et al., Extreme Ultraviolet Variability Experiment (EVE) on the Solar Dynamics Observatory (SDO): overview of science objectives, instrument design, data products, and model developments, 2012, Solar Phys., 275, 115.

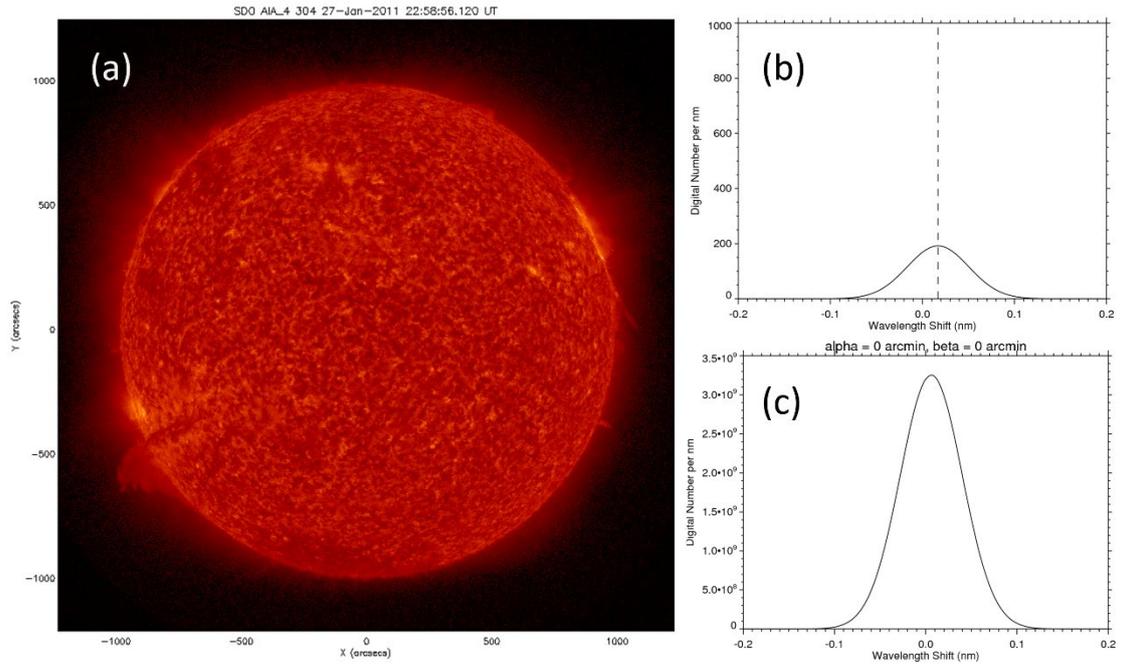

Figure 1. (a) AIA 304 image at 22:58:56 UT on 27 Jan 2011. (b) An example showing the simulated irradiance from one pixel, where the position of dashed line indicates the optical offset. (c) Full-disk irradiance profile derived from the total irradiance of all pixels.

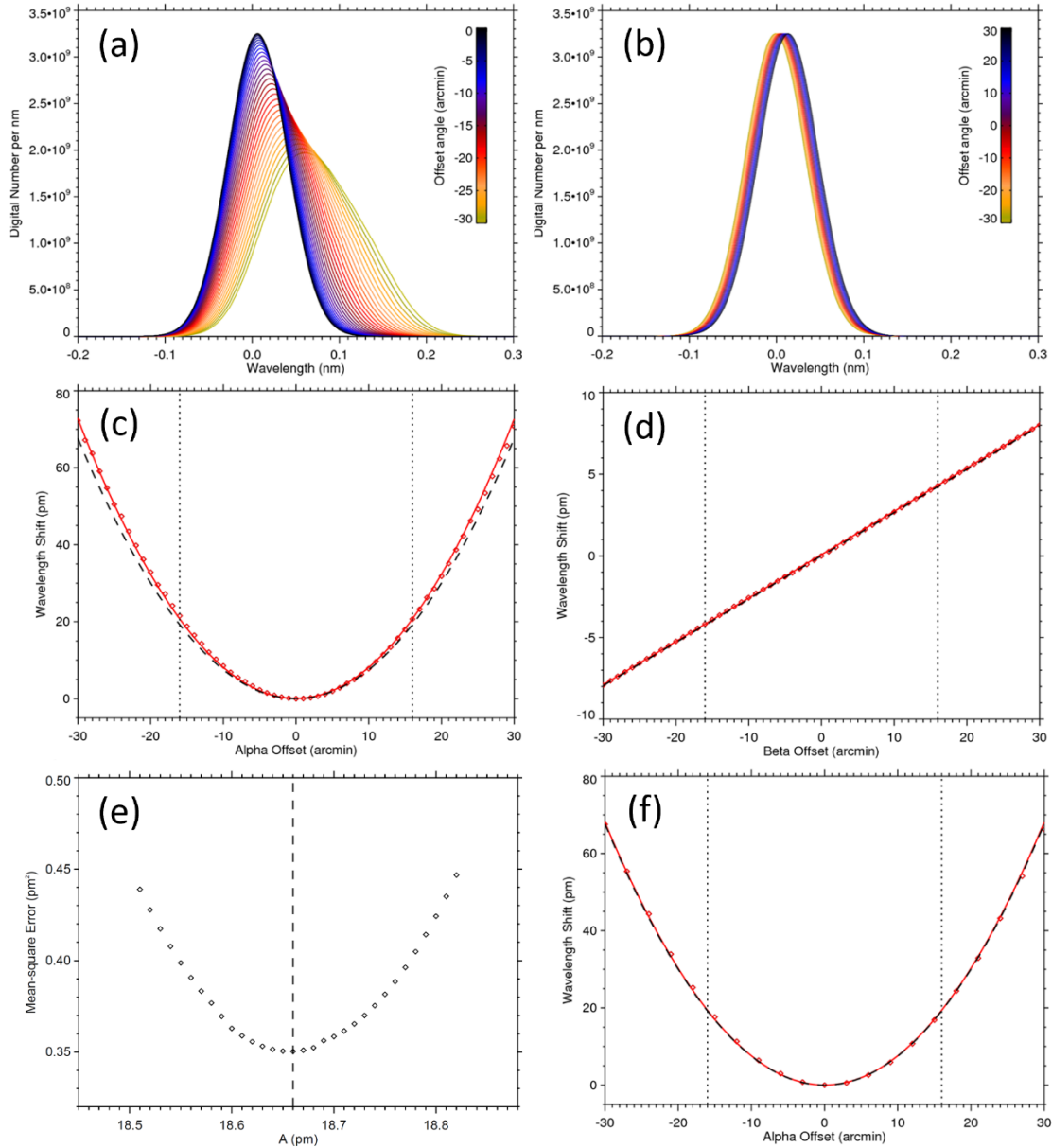

Figure 2. (a) Full-disk irradiance profiles at alpha offset angles ranging from -30' to 0', colored from yellow to blue, presenting the instrumental effect along the solar longitude. (b) Full-disk irradiance profiles at beta offset angles ranging from -30' to 30', colored from yellow to blue, presenting the instrumental effect along the solar latitude. (c) - (d) EVE's optical shift as a quadratic function of alpha offset angle and as a linear function of beta offset angle, respectively. The red symbols and solid curves represent the simulation result, and the black dashed curves are the EVE observation during the cruciform scan on 27 Jan 2011. The vertical dotted lines indicate the positions of the east and west solar limbs. (e) The mean-square error (MSE) between the correction curves of AIA simulation and EVE observation as a function of the alpha correction coefficient, A, in Equation (2). The vertical dashed line indicates the A value at 18.66, where MSE reaches the minimum. (f) EVE's optical shift as a quadratic function of alpha offset angle, after the alpha correction coefficient is changed from 19.8 to 18.66. The vertical dotted lines indicate the positions of the east and west solar limbs.

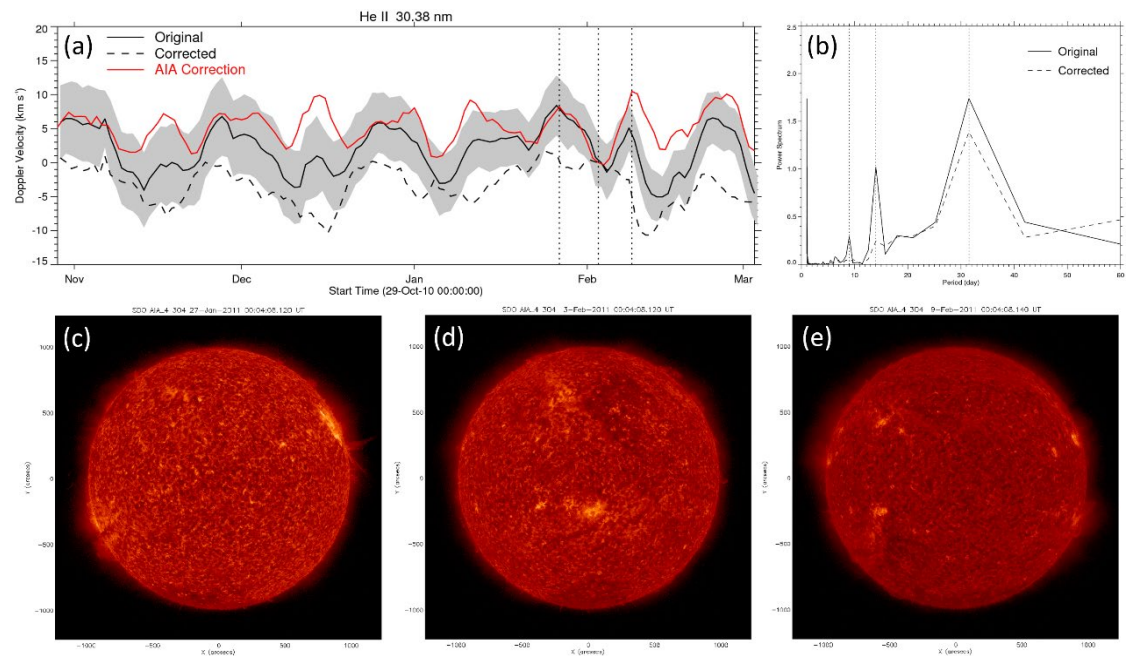

Figure 3. (a) Original and corrected He II 30.38 nm Doppler velocities from 29 Oct 2010 to 3 Mar 2011, as well as the AIA correction given by Equation (3) and AIA 304 images. The shadow indicates the 95$^{th}$ percentile distribution of the original Doppler velocity on each day. Positive/negative values correspond to the red/blue-shift. The dotted vertical lines show the times of AIA 304 images in (c) - (e). (b) Fast Fourier Transforms (FFTs) of original and corrected He II 30.38 nm Doppler velocities from 29 Oct 2010 to 3 Mar 2011. The dotted vertical lines indicate the three main periods of 9, 14 and 31.5 days. (c) - (e) AIA 304 images at 27 Jan 2011 00:04 UT, 3 Feb 2011 00:04 UT and 9 Feb 2011 00:04 UT, respectively.

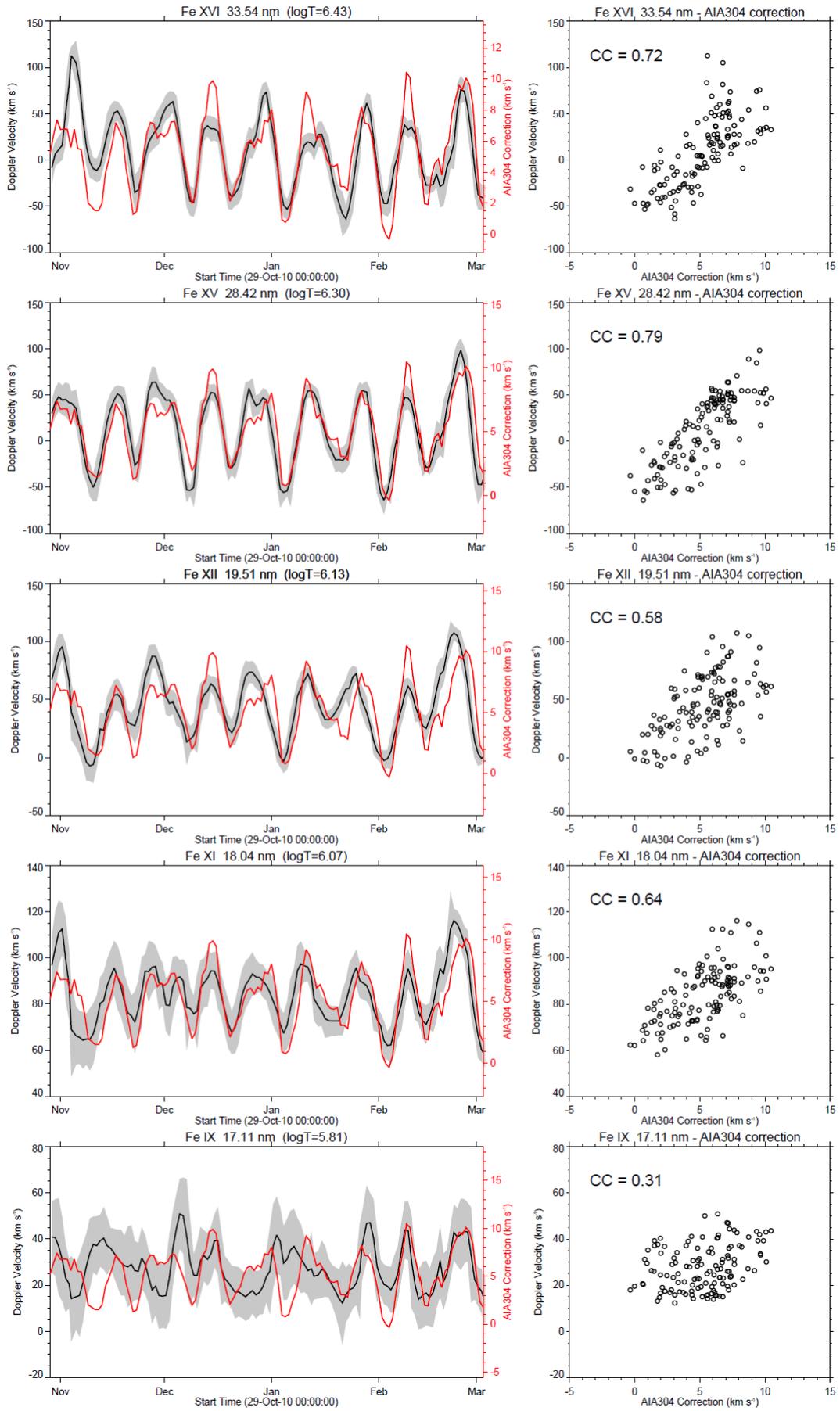

Figure 4. Left panel: Comparison between observational Doppler velocities and AIA

304 correction from 29 Oct 2010 to 3 Mar 2011 for Fe XVI 33.54 nm, Fe XV 28.42 nm, Fe XII 19.51 nm, Fe XI 18.04 nm and Fe IX 17.11 nm, respectively. Black curves and shadows indicate the Doppler velocities and their 95$^{th}$ percentile distributions, and red curves indicate the AIA304 correction. The reference wavelengths are obtained from CHIANTI atomic database. Right panel: Correlation results between their Doppler velocities and the AIA 304 correction.